\documentclass{PoS}
\usepackage{amsmath}
\usepackage{subfig}

\title{Impact of Dark Matter Direct and Indirect Detection on Simplified Dark Matter Models}

\ShortTitle{Simplified $Z^{'}$ scenarios}

\author{\speaker{Giorgio Arcadi}, Yann Mambrini and Mathias Pierre \\
\thanks{Preprint number: LPT-Orsay-15-69}\\
Laboratoire de Physique Th\'eorique  Universit\'e Paris-Sud, F-91405 Orsay, France \\
        E-mail: \email{giorgio.arcadi@th.u-psud.fr,yann.mambrini@th.u-psud.fr,mathias.pierre@th.u-psud.fr}}

\abstract{We discuss simple extensions of the Standard Model featuring a (fermionic) stable DM candidate interacting with SM fermions through a $Z^{'}$ mediator. These kind of models offer a wide phenomenology but result, at the same time, particularly manageable, given the limited number of free-parameters, and offer a broad LHC phenomenology. We will discuss the impact Direct and Indirect Dark Matter searches, assuming the latter to be thermal WIMPs. We will show in particular that the combinations of the limits on the DM Spin Independent and Spin Dependent scattering cross-section on nuclei already exclude large portions of the parameter space favored by DM relic density, in particular if, in addition, a DM Indirect signal, like the Galactic Center gamma-ray excess is required.}

\FullConference{The European Physical Society Conference on High Energy Physics\\
		22--29 July 2015\\
		Vienna, Austria}

\begin{document}

\section{Introduction}

\noindent
Weakly Interacting Massive Particles (WIMPs) are the most popular category of Dark Matter candidates. A very simple phenomenological description consists on a two particle extension of the Standard Model featuring a DM candidate directly coupled to a single state, dubbed mediator, which is in turn coupled with Standard Model fermions. These simple setups feature a rather limited set of free-parameters but, nonetheless, allow a description of the possible detection signals, ranging from Direct to Indirect Detection as well as to hypothetical collider detection both of the visible decay channels of the mediator (resonances in dilepton/dijet distributions) and decay into DM pairs (mono-object events). 
\noindent
In this work we will discuss a specific class of this simplified models featuring a fermionic (dirac) DM candidate and a spin-1, namely $Z^{'}$, mediator. 
\noindent
We will show in the next sections how the combination of constraints from DM phenomenology is effective in determining the relevant parameters of the scenario under investigation.

\section{General implementation}

\noindent
The interactions of the $Z^{'}$ with the Dark Matter candidate and the SM fermions can be schematized through the following Lagrangian: 
\begin{equation}
\mathcal{L}=\sum_f g_f \bar f \gamma^\mu \left(\epsilon^f_L P_L + \epsilon^f_R P_R\right) f Z^{'}_\mu+g_\chi \bar \chi \gamma^\mu \left(\epsilon^\chi_L P_L + \epsilon^\chi_R P_R\right) \chi Z^{'}_\mu
\end{equation} 
Featuring, as potentially free parameters, the gauge couplings $g_f$ and $g_\chi$ and the four coefficients $\epsilon^{f,\chi}_{L,R}$ associated to the left/right currents.
It is nonetheless possible, compatibly with phenomenological constraints, as discussed below, to reduce the set of free parameters by fixing the values of the couplings between the $Z^{'}$ and SM fermions. We have then considered a set of fixed assignations of the parameters $\epsilon^{f,\chi}_{L,R}$, already considered e.g. in~\cite{Han:2013mra}, summarized in tab.~(\ref{tab:Zpcouplings}).
\begin{table}[htb]
\centering
\begin{tabular}{|c|c|c|c|c|c|c|}
\hline
& $\chi$ & $\psi$ & $\eta$ & LR & B-L & SSM \\
\hline
D & $2\sqrt{10}$ & $2\sqrt{6}$ & $2\sqrt{15}$ & $\sqrt{5/3}$ & 1 & 1 \\
\hline
$\hat{\epsilon}^u_L$ & -1 & 1 & -2 & -0.109 & 1/6 & $\frac{1}{2}-\frac{2}{3}\sin^2 \theta_W$ \\
\hline 
$\hat{\epsilon}^d_L$ & -1 & 1 & -2 & -0.109 & 1/6 & $-\frac{1}{2}+\frac{1}{3}\sin^2 \theta_W$ \\
\hline
$\hat{\epsilon}^u_R$ & 1 & -1 & 2 & 0.656 & 1/6 & $-\frac{2}{3}\sin^2 \theta_W$ \\
\hline
$\hat{\epsilon}^d_R$ & -3 & -1 & -1 & -0.874 & 1/6 & $\frac{1}{3}\sin^2 \theta_W$ \\
\hline
$\hat{\epsilon}^\nu_L$ & 3 & 1 & 1 & 0.327 & -1/2 & $\frac{1}{2}$ \\
\hline
$\hat{\epsilon}^l_L$ & 3 & 1 & 1 & 0.327 & -1/2 & $-\frac{1}{2}+\sin^2 \theta_W$ \\
\hline
$\hat{\epsilon}^e_R$ & 1 & -1 & 2 & -0.438 & -1/2 & $\sin^2 \theta_W$ \\
\hline
\end{tabular}
\caption{\footnotesize{Summary table of the couplings of the $Z^{'}$ with SM fermions considered in this work. Notice that $\epsilon_{L,R}^f=\hat{\epsilon}_{L,R}^f/D$.}}
\label{tab:Zpcouplings}
\end{table}
The first three realizations in tab.~(\ref{tab:Zpcouplings}), labeled as $\chi,\psi$ and $\eta$, refer to three $E6$ inspired theoretical setups (see e.g.~\cite{Langacker:2008yv}). In addition we have considered assignations according the left-right and $B-L$ extensions of the SM gauge group, which can also be embedded in GUT frameworks based on $SO(10)$, and finally the very popular benchmark dubbed Sequential Standard Model (SSM) consisting into a $Z^{'}$ coupled to the SM fermions as the $Z$ boson. The couplings of the $Z^{'}$ with the DM will be instead regarded as free.
\noindent
For phenomenological reasons it is more convenient to introduce vectorial and axial couplings according the following parametrization:
\begin{align}
& g_2 V_{f}=\frac{g_f}{2} \left(\epsilon^f_{L}+\epsilon^f_{R}\right),\,\,\,\,\,g_2 A_{f}=\frac{g_f}{2} \left(\epsilon^f_{L}-\epsilon^f_{R}\right)\nonumber\\
& g_2 V_\chi=\frac{g_\chi}{2} \left(\epsilon^\chi_{L}+\epsilon^f_{R}\right),\,\,\,\,\,g_2 A_{\chi}=\frac{g_\chi}{2} \left(\epsilon^\chi_{L}-\epsilon^\chi_{R}\right)
\end{align}
where $g_2=g$ for the SSM and $g_2=\sqrt{\frac{5}{3}}g \tan\theta_W \sim 0.46$ for the other models.

\noindent
This parametrization reduces the number of free parameters of the theory to four, i.e. the masses $m_\chi$ and $m_{Z^{'}}$ and the vectorial and axial couplings $V_\chi$ and $A_\chi$. 

\section{Correlation between dark matter constraints}

\noindent
In this section we show the effectiveness of the correlation among the different constraints from dark matter phenomenology, namely dark matter Direct and Indirect detection as well as the correct relic density according the WIMP paradigm. Such correlation can be understood in terms of a relation of the type: 
\begin{equation}
\label{eq:schematic_relation}
\langle \sigma v \rangle = f_1 \sigma_{\chi N}^{\rm SI}+ f_2 \sigma_{\chi N}^{\rm SD}
\end{equation}
between the DM annihilation cross-section, responsible of the relic density and, possibly, and Indirect Detection signal, and the two conventional components, Spin Independent (SI) and Spin Dependent (SD), of the DM scattering cross-section on nuclei.
\noindent
This kind of relation has a twofold implication. On one side it is possible to use constraints from current DM Direct Detection experiments to test the WIMP hypothesis. On the other side it is possible to predict hypothetical future signals at Direct Detection facilities by requiring that the DM features the correct thermal relic density.

\noindent
The SI and SD cross-sections (for definiteness we will report the cross-section on protons) can be straightforwardly related to, respectively, the couplings $V_\chi$ and $A_\chi$ as:
\begin{equation}
\sigma_{\chi \,p}^{SI}=\frac{g_2^4 |V_\chi|^2}{\pi m_{Z^{'}}^4} \alpha_{\rm SI},\,\,\,\,\,\,\,\,\sigma_{\chi \,p}^{SD}=\frac{3 g_2^4 |A_\chi|^2}{\pi m_{Z^{'}}^4} \alpha_{\rm SD}
\end{equation}

\noindent
where the factors $\alpha_{\rm SI}$ and $\alpha_{SD}$, which depend on $V_f$ (SI) and $A_f$ (SD) as well as on the target nuclei (see e.g.~\cite{Arcadi:2013qia} for the full expressions) allow to compare theoretically predicted cross-sections, typically characterized by different interactions of the DM with protons and neutrons, with experimental limits, customarily casted under the assumption of same interactions for protons and neutrons.  The $Z^{'}$ realizations proposed in tab.~(\ref{tab:Zpcouplings}) feature both non negligible SI and SD cross sections ad exception of the $\psi$ model, missing the SI component since $V_f=0$, and the $B-L$ model, missing the SD component since $A_f=0$~\footnote{In reality the missing component are originated by 'crossed' combinations like e.g. $V_f A_\chi$ (see~\cite{DelNobile:2013sia} for a review). However in these cases the cross-section are suppressed by powers the momenta of the incoming dark matter particle and result much below current experimental sensitivity. As a a consequence they will be neglected in our analysis.}.

\noindent
The DM annihilation cross-section can be as well casted in terms of the couplings $V_\chi$ and $A_\chi$. A particularly simple expression is obtained, away from s-channel resonances, once the conventional velocity expansion is performed:
\begin{align}
& \langle \sigma v \rangle_{f \bar f} \simeq \frac{g_2^4 m_\chi^2}{2 \pi m_{Z^{'}}^4} \sum_f n_c^f \left( |V_f|^2 + |A_f|^2  \right)\nonumber\\
&|V_\chi|^2 \left[ 2  + \alpha^2 \left( \frac{m_b^2}{m_\chi^2}  \frac{|A_b|^2}{\sum_f ( |V_f|^2 + |A_f|^2 ) }+\frac{m_t^2}{m_\chi^2}  \frac{|A_t|^2}{\sum_f ( |V_f|^2 + |A_f|^2 ) } + \frac{v^2}{6} \right) 
\right]\nonumber\\
& = \frac{m_\chi^2}{2 \mu_{\chi p}^2} \sum_f n_c^f \left( |V_f|^2 + |A_f|^2  \right)\nonumber\\
& \left[2 \frac{\sigma_{\chi p}^{\rm SI}}{\alpha_{\rm SI}}+\frac{\sigma_{\chi p}^{\rm SD}}{3 \alpha_{\rm SD}}\left( \frac{m_b^2}{m_\chi^2}  \frac{|A_b|^2}{\sum_f ( |V_f|^2 + |A_f|^2 ) }+\frac{m_t^2}{m_\chi^2}  \frac{|A_t|^2}{\sum_f ( |V_f|^2 + |A_f|^2 ) } + \frac{v^2}{6} \right)\right] 
\label{Eq:sigvff}
\end{align}
where $\alpha=\frac{|A_\chi|}{|V_\chi|}$ and $n_c^f$ is the color number.

\noindent
Interestingly, in the simple limit considered, the ratio between the dark matter annihilation cross-section and the SI/SD scattering cross-sections is independent from the mass of the $Z^{'}$. It is then possible, by computing the ratio between the annihilation cross-section and the SI cross-section, which is strongly constrained by LUX experiment~\cite{Akerib:2013tjd}, to determine, for each $Z^{'}$ realization, the value of $\alpha$, as function of the DM mass, compatible with the correct relic density, as obtained by imposing the value $\langle \sigma v \rangle \approx 3 \times 10^{-26} {\mbox{cm}}^3\,{\mbox{s}}^{-1}$:
\begin{equation}
\alpha \approx 1.1\times 10^{3}\sqrt{\frac{\alpha_{\rm SI}}{\sum_f n_c^f |V_f|^2+|A_f|^2}} {\left(\frac{\langle \sigma v \rangle}{3 \times 10^{-26}\,{\mbox{cm}}^{3} {\mbox{s}}^{-1}}\right)}^{1/2}{\left(\frac{\sigma_{N\chi}^{\rm SI}}{10^{-44}\,{\mbox{cm}}^{2}}\right)}^{-1/2}{\left(\frac{m_\chi}{100\,\mbox{GeV}}\right)}^{-1}
\end{equation}

\noindent
As evident, due to the strong constraints on the SI cross-section, the WIMP paradigm seems to favor axial-type couplings between the DM and the $Z^{'}$. We can then determine a prediction of the SD cross-section as function of $\langle \sigma v \rangle$:
\begin{equation}
\label{eq:SD_prediction}
\sigma_{N\chi}^{\rm SD} \approx 1.6 \times 10^{-37}\,{\mbox{cm}}^2 \frac{\alpha_{\rm SD}}{n_c^f \sum_f |V_f|^2+|A_f^2|} \left(\frac{\langle \sigma v \rangle}{3 \times 10^{-26}\,{\mbox{cm}}^{3} {\mbox{s}}^{-1}}\right){\left(\frac{m_\chi}{100\,\mbox{GeV}}\right)}^{-2}  
\end{equation}
which can be probed by current or, possibly, next future experiments.

\begin{figure}[htb]
\begin{center}
\subfloat{\includegraphics[width=6.5 cm]{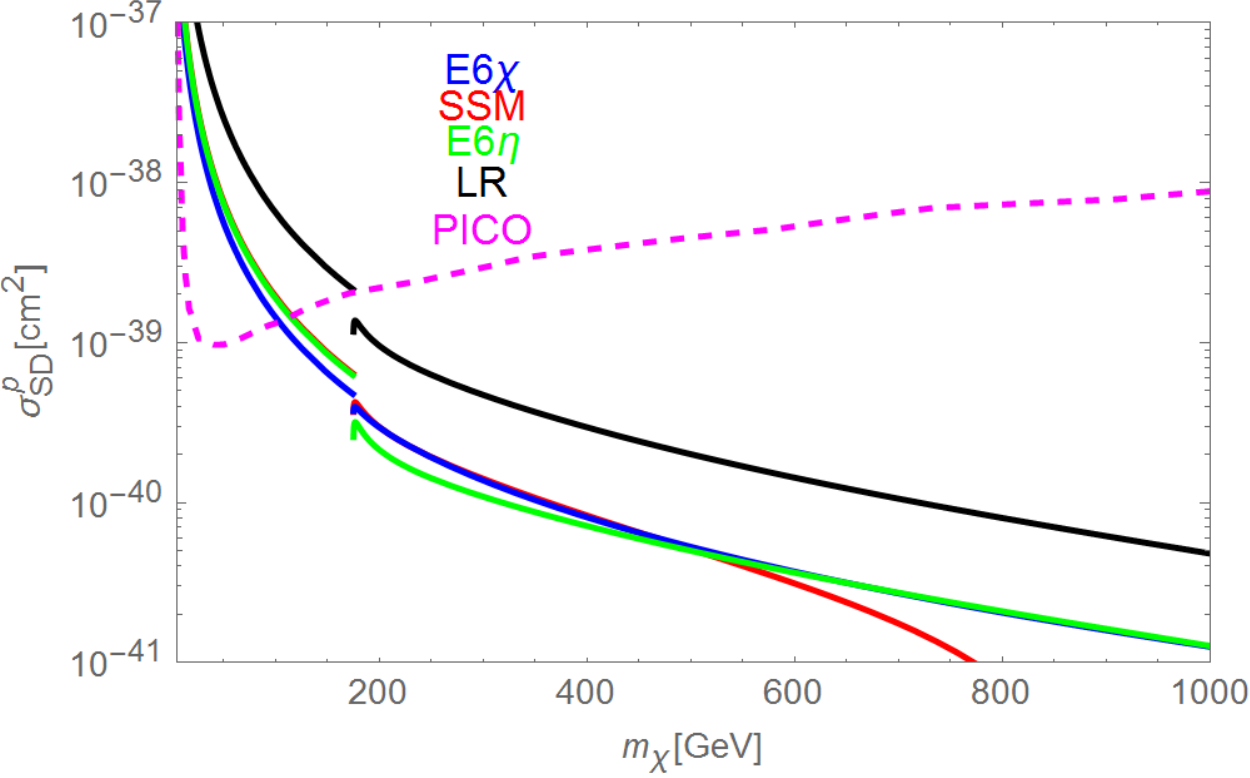}}
\subfloat{\includegraphics[width=6.5 cm]{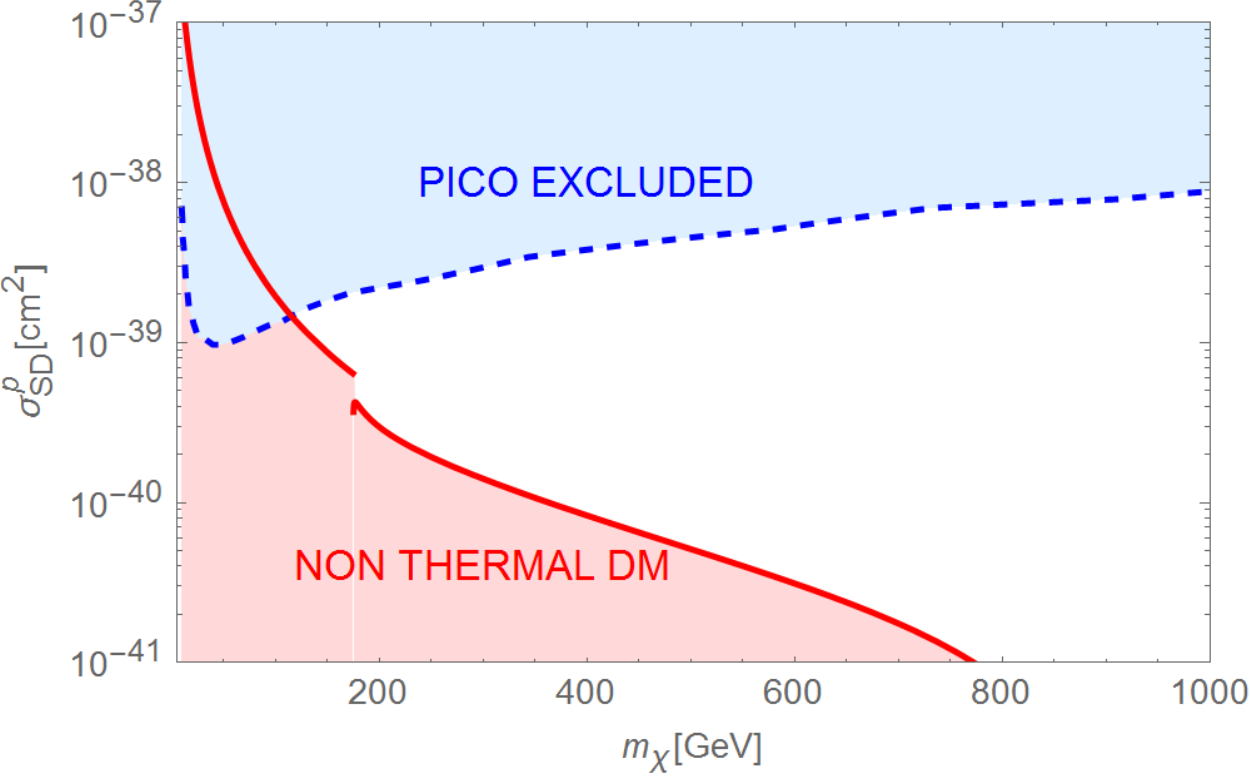}}
\end{center}
\caption{\footnotesize{Left panel: Values of the predicted SD scattering cross-section on protons by imposing the correct DM relic density for the considered $Z^{'}$realizations. These values are confronted with the experimental limit (magenta dashed lines) from PICO (proton). For each model the regions above the magenta curves are excluded by DM direct detection while below the solid lines only non-thermal production of DM is viable. Right panel: The same as left panel but highlighting the excluded regions. For more clearness we have focused on just one $Z^{'}$ realization, namely the SSM.}}
\label{fig:alphas}
\end{figure}

\noindent
As an example we have compared, in fig.~(\ref{fig:alphas}), eq.~(\ref{eq:SD_prediction}) with the limit on the SD cross-section on proton as recently given by the PICO collaboration~\cite{Amole:2015lsj}, being the most stringent up to now. Remarkably, current constraints are already capable of probing the thermal dark matter hypothesis. For the $Z^{'}$ realizations considered, thermal dark matter is excluded for masses below $100-150\,\mbox{GeV}$.

\noindent
The result discussed above can be also applied to determine possible prospects for Dark Matter Indirect detection by mean, again, of eq.~\ref{eq:schematic_relation} in which, this time, the left hand side represents the value of the cross-section responsible of hypothetical indirect signals. It is interesting, in particular, to investigate whether the framework under consideration can account for the Galactic Center (GC) gamma ray excess~\cite{GC}, possibly originated by a DM candidate with mass of approximately 50 GeV annihilating into $b \bar b$ with a cross-section of the order of thermal one. This last requirement implies a DM annihilation cross-section approximately constant between the time of freeze-out and present time and is accomplished only if it is dominated by the s-wave term, independent on the velocity.

\noindent
As evident from the second line of eq.~(\ref{Eq:sigvff}) a sizable s-wave contribution is provided only by the vectorial coupling $V_\chi$, being the one associated to $A_\chi$ chirality suppressed. However, in order to comply with limits from the SI cross-section which require $V_\chi \ll 1$, one has to rely on the axial contribution for achieving the correct relic density. As consequence of the mentioned chirality suppression, the DM annihilation cross-section is velocity dependent and a strong mismatch between its value at thermal freeze-out, corresponding to $v \sim 0.3$, and the present time value, corresponding to $v\rightarrow 0$, arises. The only possible exception is constituted by the $E6_\psi$ realization for which $V_\chi$ is substantially unconstrained since $V_f=0$ and, then, no sizable SI cross section is induced.

\begin{figure}[htb]
\begin{center}
\subfloat{\includegraphics[width=6.5 cm]{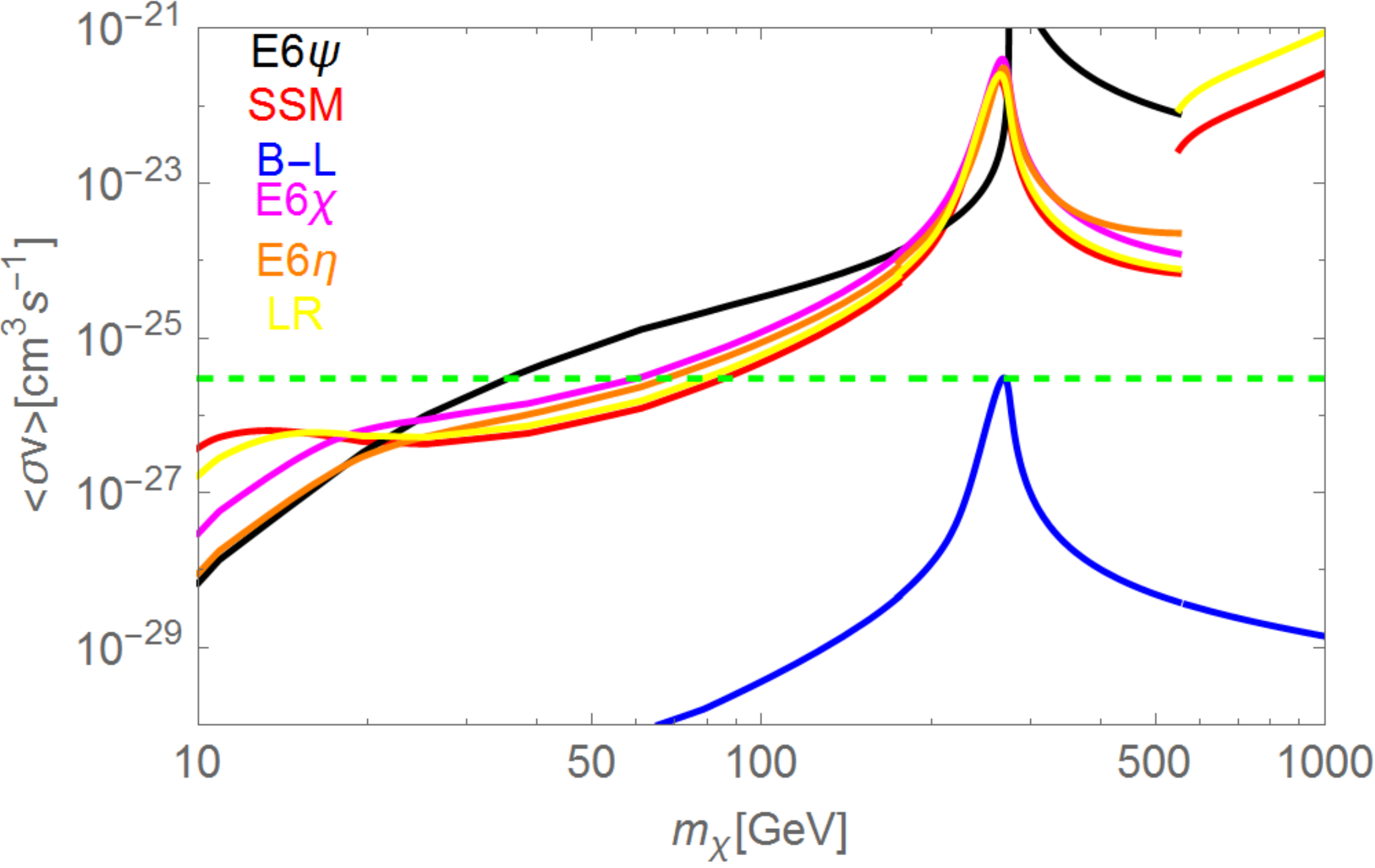}}
\subfloat{\includegraphics[width=6.5 cm]{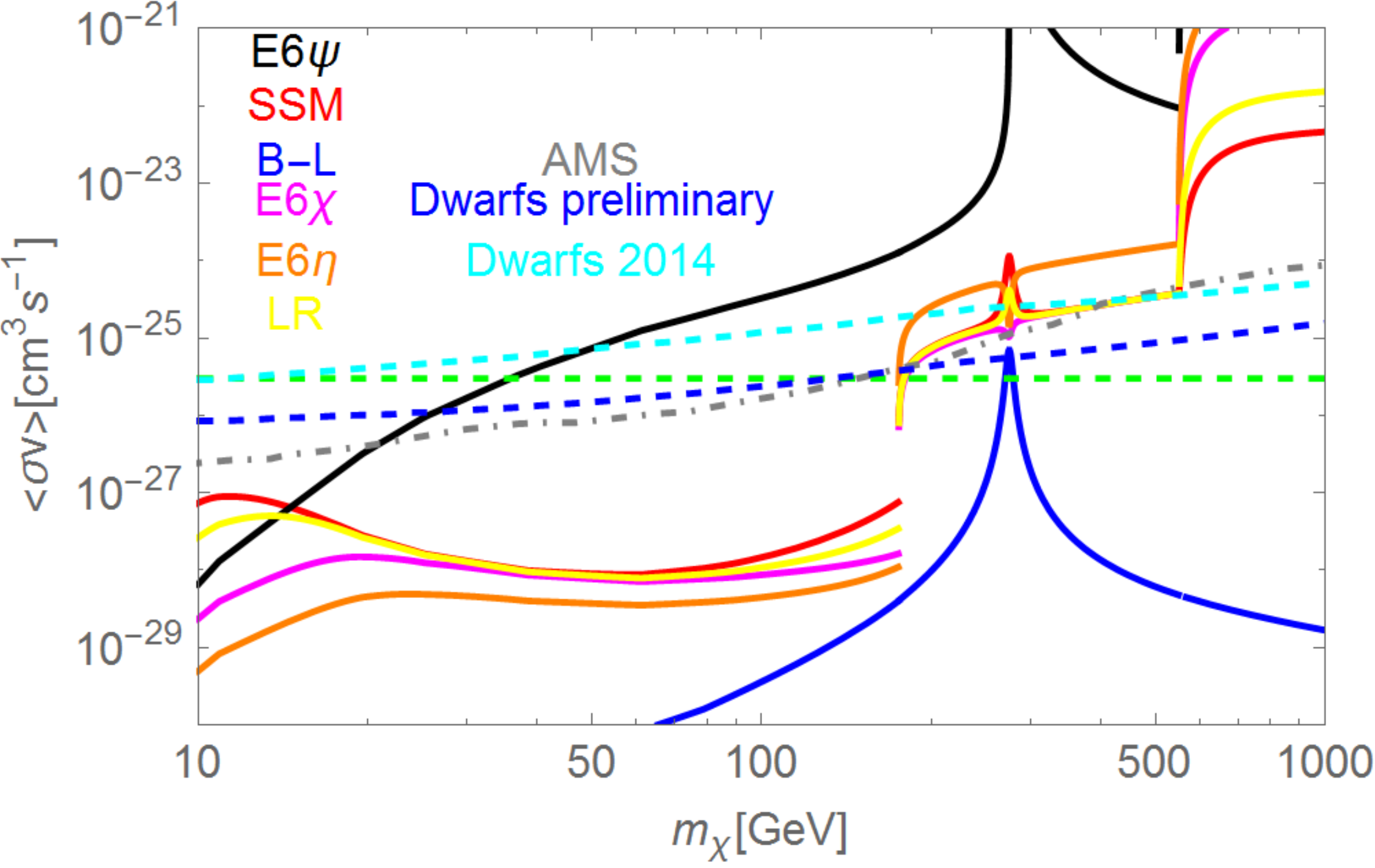}}
\end{center}
\caption{\footnotesize{Left panel: DM matter annihilation cross-section, as function of the DM mass, at the time of DM freeze-out for the considered $Z^{'}$ realizations. The couplings $V_\chi$ and $A_\chi$ have been fixed to their corresponding limits from DM DD. Right panel: The same as left panel but considering, instead, the annihilation cross-section at present times. The dashed lines represent current limits from Dsph by Fermi (cyan), their next future projection (blue) and limits by antiprotons (gray).}}
\label{fig:sigmas}
\end{figure}

\noindent
We have shown in fig.~(\ref{fig:sigmas}) two illustrative plots reporting the annihilation cross-section at freeze-out (left panel) and at present times (right-panel) as function of the DM mass and for couplings fixed with respect to the corresponding limits from Direct Detection. For the $B-L$ and $E6_\psi$ models, featuring DM couplings not strongly constrained by DD, we have set $A_{\chi,B-L}=0$ and $V_{\chi,E6_\psi}=A_{\chi,E6_\psi}$. For simplicity, a common value of $m_{Z^{'}}=550$~\footnote{\noindent For some of the realizations considered this value is already excluded by collider searches. Fig.~(\ref{fig:sigmas}) should be intended as an illustration. Our conclusion do not sensitively change by considering more realistic values of $m_{Z^{'}}$.} has been assumed. Values of the thermal cross-section compatible with viable relic density normally correspond to very suppressed present time cross-section, below current experimental sensitivity. The only exceptions occur in the case of s-channel resonances, i.e. $m_\chi \sim m_{Z^{'}}/2$, and for $m_\chi > m_{Z^{'}}$ when the additional annihilation channel $\bar \chi \chi \rightarrow Z^{'} Z^{'}$ is open. The discussion of the high DM mass regime is not in the purpose of this work and will be left to further study. The only realization featuring a sizable s-wave annihilation cross-section is the $E6_\psi$ due to the absence of constraints on the SI cross-section. In this case the annihilation cross-section can largely exceed, even at present times, the thermal value. In this kind of scenario effective complementary constraints to the ones from PICO come from Indirect Detection limits, namely gamma rays from Dsph~\cite{Ackermann:2013yva} and AMS antiprotons~\cite{Giesen:2015ufa}.

\section{Conclusions}

\noindent
Simplified DM models are a very simple but, at the same time, powerfull tool for interpreting the outcome of Dark Matter searches and, more in general, searches of New Physics. We have discussed in this work a specific example, consisting into and extension of the Standard Model with a Dirac Dark Matter candidate and a $Z^{'}$ mediator. We have shown the effectiveness of the correlation among different Dark Matter searches and the requirement of the correct relic density. Direct Detection favors DM axially coupled with the $Z^{'}$ unless particular assignations of the couplings with SM fermions, like e.g. in the $E6_\psi$ realization, are considered. In this last case complementary constraints are provided by Dark Matter Indirect Detection.

\noindent
{\bf Acknowledgements}:G.A. acknowledges support from the ERC advanced grant Higgs@LHC.

\end{document}